\documentclass{statsoc}
\usepackage[a4paper]{geometry}
\usepackage{graphicx, array, blindtext}
\usepackage{amsmath}
\usepackage{natbib}
 \usepackage{booktabs}
\usepackage{bbm}

\usepackage{hyperref}
\usepackage{float}
\usepackage{etoolbox}

\makeatletter
\patchcmd{\@makecaption}
{\parbox}
{\advance\@tempdima-\fontdimen2} % decrease the width!
{}{}
\makeatother

\title[Stochastic Gradient Relational Event Additive Model]{A Stochastic Gradient Relational Event Additive Model for modelling US patent citations from 1976 until 2022}
\author[Filippi-Mazzola and Wit]{Edoardo Filippi-Mazzola}
\address{Institute of Computing, Università della Svizzera italiana,
Lugano,
Switzerland.}
\email{edoardo.filippi-mazzola@usi.ch}
\author[Filippi-Mazzola and Wit]{Ernst C. Wit}
\address{Institute of Computing, Università della Svizzera italiana,
Lugano,
Switzerland.}

\begin{document}
\begin{abstract}
%The patent citation network is a complex and dynamic system that reflects the diffusion of knowledge and innovation across different fields of technology. We focus on US patent citations from 1976 until 2022, which involves almost 10 million patents and over 100 million citations. Analyses of such networks are often limited to descriptive statistics. 
%Instead, in this work we aim to develop a generative model for the citation process by combining relational event models (REMs) and machine learning techniques. We propose a stochastic gradient relational event additive model (STREAM) that models the relationships between cited and citing patents as events that occur over time, capturing the dynamic nature of the patent citation network. Each predictor in the generative model is assumed to have a non-linear behaviour, which has been modeled via a B-spline. By estimating the model through an efficient stochastic gradient descent approach, we are able to identify the key factors that drive the patent citation process. 
%Our analysis revealed several interesting insights, such as the identification of time windows in which citations are more likely to happen, and the relevancy of the increasing number of citations received per patent. Overall, our results demonstrate the potential of the STREAM in capturing complex dynamics that arise in a large sparse network, maintaining the features and the interpretability, for which REMs are most famous.  

Until 2022, the US patent citation network contained almost 10 million patents and over 100 million citations. To overcome limitations in analyzing such complex networks, we propose a stochastic gradient relational event additive model (STREAM) that models the relationships between citing patents as events that occur over time, where predictors are modeled through B-splines. Our model identifies key factors driving patent citation and reveals insights, such as time windows where citations are more likely and the relevance of the increasing citation numbers per patent. Overall, the STREAM offers the potential for capturing dynamics in large sparse networks while maintaining interpretability.
\end{abstract}


\section{Introduction}

%Analyzing the patent citation network is a crucial aspect of understanding the dynamics of innovation in various industries. 
Patents are not only a means of protecting intellectual property but also provide valuable information about the state of the art in technology and the evolution of knowledge and innovation over time \citep{jaffe_innovation_2002}. The patent citation network captures the relationships between patents, where each citation represents a connection between two patents, indicating that the citing patent has built upon the knowledge contained in the cited patent \citep{sharma_knowledgeflow_2017}. 
%By analyzing the patent citation network, we can gain insights into the diffusion of knowledge and develop a deeper understanding of the patterns and trends that drive innovation. Furthermore, 
Patents represent a significant investment for many companies, and understanding the competitive landscape, and the strengths and weaknesses of competitors' patent portfolios can be essential for making strategic decisions about technology development, licensing, and litigation \citep{lerner_importanceinvestnemtns_1994}. Analyzing the factors that lead to a patent being cited can provide valuable insights into the underlying mechanisms driving innovation. Additionally, understanding the drivers of patent citation can inform decision-making in a variety of contexts, such as technology development, intellectual property management, and innovation policy \citep{holger_ipmanagment_2003}. However, patent data analysis is a complex and challenging task, requiring advanced techniques and tools for managing and analyzing large and complex datasets. 

The relational event model (REM) \citep{butts_2008,perry_point_2013} has emerged as a powerful tool for modelling complex relational data. Although REMs were first introduced in the social sciences as a way of modelling the temporal dependencies between interactions in social networks, they have been applied in many different contexts, such as two-mode networks \citep{vu_relational_2017}, animal behavioural interactions \citep{tranmer_animal_2015}, and more recently, financial transactions \cite{bianchi_clocks_2022}. Following these examples, REMs can be a valuable tool for analyzing  citation networks of patents, as they allow us to model the complex relationships between citing and cited patents, identifying the factors that influence the diffusion of knowledge and innovation. However, the practical applicability of REMs is limited by their runtime complexity \citep{welles_comunication_2014}, a problem rooted in the denominator of the partial likelihood on which estimation of most REMs is based.  Although there have been some early attempts to model citation networks through a REM-like approach \citep{vu_dynamic_2011}, only recently \cite{lerner_reliability_2020} tackled the inherent computational issues by showing the robustness of REM estimation when controls and cases are sub-sampled through a nested case-control approach \citep{borgan_methods_1995}. This was first introduced in REMs by \cite{vu_relational_2015}. 

Nevertheless, the tools to apply these models to big data still show practical limitations in memory management and optimization procedures. Stochastic gradient relational event additive models solve the memory management and optimization issues using concepts from the machine learning field. It uses  an Adaptive Moment (ADAM) optimizer \citep{kingma_adam_2017} to improve the efficiency of coefficient estimation. Furthermore, it improves on existing approaches by incorporating the use of B-Splines \citep{Schoenberg1946Contributions, Schoenberg1969interpolation, DeBoor1972}. By modelling each predictor through a B-spline, STREAM can capture non-linear relationships between variables, providing more valuable interpretations to time-varying effects while identifying the most influential factors driving patent citation. 

%Meanwhile, the use of the ADAM optimizer allows for faster and more efficient estimation of coefficients, making STREAM an excellent choice for large-scale applications.

%Building on the foundation of REMs, we present in this paper a Stochastic Gradient Relational Event Additive Model (STREAM) as an alternative to model large-scale complex event sequences in a REM-like fashion. 

%This novel modelling technique incorporates the use of B-Splines \citep{Schoenberg1946Contributions, Schoenberg1969interpolation, DeBoor1972} and the Adaptive Moment (ADAM) \citep{kingma_adam_2017} optimizer to improve coefficient estimation and modelling flexibility. By modelling each predictor through a B-spline, STREAM can capture non-linear relationships between variables, providing more valuable interpretations to time-varying effects while identifying the most influential factors driving patent citation. Meanwhile, the use of the ADAM optimizer allows for faster and more efficient estimation of coefficients, making STREAM an excellent choice for large-scale applications.

For our analysis, we used data obtained from the United States Patent and Trademark Office (USPTO), the federal agency responsible for granting patents and registering trademarks in the United States. The USPTO data is one of the most comprehensive sources of patent information in the world as it contains precise information contained in standard digitalized formats on all patents issued in the United States since 1976. While there are limitations to extrapolating the USPTO data to other regions, it is still a good proxy for global patent activity as well as source for studying innovation and technological progress. Overall, by using the STREAM model, we gained important insights into the dynamics of patent citations while opening the road to further speculations on how the current state of the innovation process.

In this paper, we start by describing the USPTO patent data in section~\ref{patent_citations} on which this analysis is based. After developing STREAM's theoretical foundation in section~\ref{STREAM}, we apply the framework to the patent citation network in section~\ref{effects}. Although the STREAM was specifically designed to work with citation networks, this modelling framework can easily be applied to model general relational event data. 

%Indeed, this legal requirement poses the question on what exactly mean to be related to a specific techonoogy on which legal protection is asked. By now, it is clear from previous studies \citep{kuhn_information_2010,kuhn_patent_2020,filippi_similarity_2022} that the patent citaiton process has gone through some fundamental evolution that marked a change in the data generation process. 

\section{Patent citations as event history data}\label{patent_citations}

Patent citation is an essential element of the patent system as it provides a means of demonstrating the novelty, non-obviousness, and importance of an invention. Indeed, a patent citation is crucial for both patent examiners and inventors, as they allow the examiner to evaluate the claims made in the patent application, and it helps the inventor establish the scope and value of their invention. In this regard, in many jurisdictions, applicants are legally obliged to cite those patents on which the patent builds forth as part of a patent deposition. The triple consisting of the instance of deposition, the citing, and cited patents can be seen as an instance of a relational event. Collections of patent citations constitute a citation network, which is a particular kind of temporal-directed graph, where new actors join the network and bind to existing nodes. In most situations, the citation is due to content similarity or other exogenous drivers. This is in contrast to classic social network architectures, where tie formation is a more endogenous process, based on, e.g., repetition, reciprocity, or triadic effects. 

In large jurisdictions, patent citation networks consists of millions of time-stamped recorded citation events. The generative process of the US patent deposition gives important clues for modelling the resulting citation network. When a patent is filed, the owners have a legal requirement to fulfill the duty of disclosure. This consists of providing within the application a list of existing technologies or scientific discoveries that are related or considered to be fundamental for the creation of the patenting invention. Patent office examiners will only grant  the patent if the application meets the uniqueness requirement and if the invention is fully disclosed in the documentation presented. The patent citation process conforms to the specific structure of event history data. The event set consists of a citation-based relationship between a specific sending, deposited patent and a receiving, pre-issued patent.  

The patent citation network suffers from several boundary issues, relating to both space and time. With regard to space, different national or trans-national jurisdictions have different application processes. Despite their similarities, slight differences in the juridical procedures make the citation generating process both country- or region-specific. A clear example of this is the following difference between the citation procedures between the European Patent Office (EPO) and the United States Patent and Trademark Office (USPTO). In the latter, it is the duty of the examiner committee to integrate additional documents and patent citations. EPO examiners, on the other hand, do not include any citations but evaluate if the invention has been properly disclosed by the cited documents. This difference  results in USPTO patents citations typically surpassing the EPO patent citations by a large amount, sometimes referred to as a ``patent office bias'' \citep{bacchiocchi_international_2010}. We focus in this study on the USPTO patent citation network. 

With respect to the time boundary, the electronic recording of patent citations has only started in relatively recently. Although some sporadic efforts have been undertaken to record historic patent citations, this is far from complete.  
We focus our analysis on those patents issued by the USPTO between 1976 and 2022. The starting year of our observed period coincides with the initialization of the digitization process of US-patents. 

In our analysis we make use of the original USPTO online repository (\url{https://bulkdata.uspto.gov/}). This makes the raw materal of this analysis as much standardized as possible in terms of general information available. Although there are various distributions available of the USPTO data, after careful evaluation we decided to avoid any third party pre-processing. The raw USPTO XML files were processed in a uniform manner and combined to obtain CSV files through an open-source software available at \url{https://github.com/efm95/patents}.

The resulting USPTO patent citation dataset consists of more than 8 million issued patents that generated 190 million citations. Despite the in-house processing by the USPTO, we have applied some data cleaning procedures as a result of some specific features of the USPTO patents. First, by focusing our view on patents issued only by the USPTO means losing track of those citations that go to patents outside the US jurisdictions. Secondly, in the same way as \cite{whalen_patent_2020} and \cite{filippi-mazzola_drivers_2023}, we excluded all non-utility patents, such as plant and design patents, as these differ in many structural ways from the utility patents. With these two additional steps, our final dataset consists of around 100 million citations issued by a network of 8.3 million patents.

\begin{figure}[tb]
	\centering
	\includegraphics[width=0.7\textwidth]{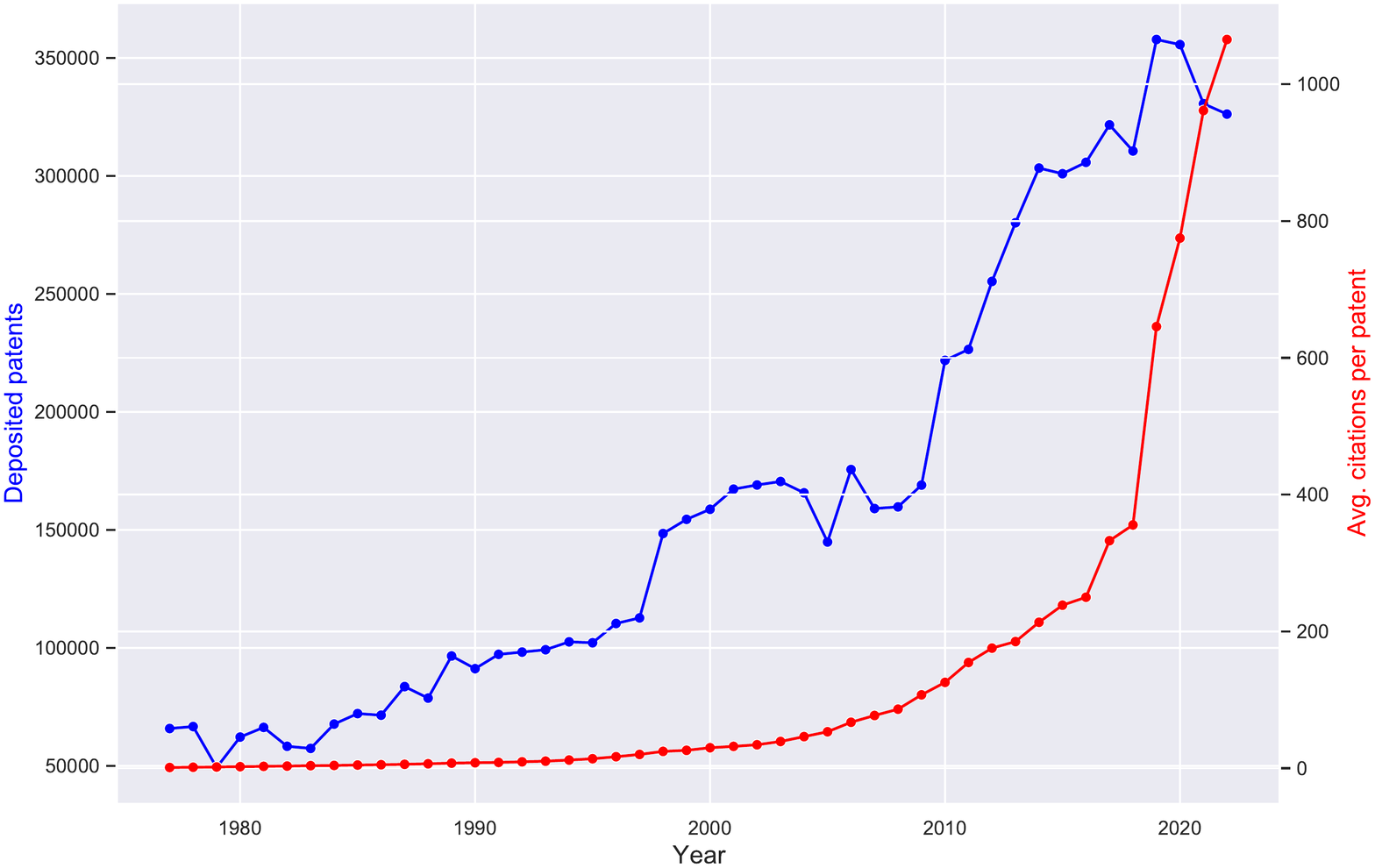}
	\caption{\label{fig:deposited-vs-cit}
	Number of deposited patents per year and the number of patent citations per patent per year since 1976.}
\end{figure}

Figure~\ref{fig:deposited-vs-cit} shows that there has been a steady increase since 1976 in the number of deposited patents per year, and a dramatic rise in the number of citations per patent. It is clear that various regulatory considerations have played a role. Failing to take those aspects into account will confound the picture of the true underlying innovation process.  The aim of this paper is to disentangle the causes that have contributed to this rise. 

\section{Stochastic Gradient Relational Event Additive Model (STREAM)}\label{STREAM}

Relational event models (REMs) are a class of statistical models used to analyze event sequences and relationships between actors through a series of exogenous and endogenous effects based on the fine-grained event history process. In this section, we will extend the REM by developing the stochastic gradient relational event additive model for the network of patent citations.

\subsection{Relational event model}

The temporal dynamic network of patent citations is represented by a sequence of time-stamped events. Each citation event $e_i$, for $i = 1,\dots,n$, is recorded as the triple $e_i=(s_i,r_i,t_i)$, where $s_i$ is the citing patent (sender), $r_i$ the cited patent (receiver) and $t_i$ the time at  which the event takes place, corresponding to the publication date of the citing patent. As in \cite{perry_point_2013}, we define a counting process for the directed event that involves sender patent $s$ and receiver patent $r$ as 
\begin{equation*}
	N_{sr}(t)=\#\{ s \text{ cited } r \text{ up to time } t\}.
\end{equation*}
The process is trivially zero up to the publication time $t_s$ of $s$, and only becomes one if patent $s$ cites patent $r$ at the time of its publication. The counting process $N_{sr}(t)$ is a local submartingale for which it is possible to define a predictable increasing process $\Lambda_{sr}(t)$, whose stochastic intensity function $\lambda_{sr}(t_s)$ describes the tendency for $s$ to cite $r$ at the time of its publication $t_s$. Given the history of the network $\mathcal{H}_{t^-}$ up to time $t$, it is possible to model the intensity function following the proportional hazard function \citep{cox_regression_1972}. The intensity function is given as the product of a baseline hazard $\lambda_0$ and an exponential function of $q$ covariates $x_{sr}(t)$ with corresponding parameter $\beta$, i.e.,
\begin{equation}\label{eq:rem}
	\lambda_{sr}(t \mid \mathcal{H}_{t^-})=  \lambda_0(t)  e^{\sum_{k=1}^{q} \beta_k  x_{srk}(t )}.
\end{equation}
While events are assumed to be conditionally independent given the network of previous events, the inclusion of covariates in this model specification allows to examine the impact of various drivers related to senders, receivers, or network topology. 

Given the difficulties that come with dealing with the full likelihood in \eqref{eq:rem}, it is possible to estimate the coefficients through a partial likelihood approach \citep{cox_partial_1975}, in which the baseline is treated as a nuisance parameter. The main idea of this approximation is to specify a partial likelihood that depends only on the order in which events occur, not the times at which they occur. Because the event time is by definition the publication date of the sender, the risk-set $R(t \mid \mathcal{H}_{t^-})$ consists of all potential receivers $r$ that were present in the network at time $t$ and, as a consequence, that could have been cited by the issued patent $s$. This results in the following partial likelihood,
\begin{equation}\label{eq:pl}
	PL(\beta) = \prod_{i = 1}^n\left( \frac{\exp\left\{\sum_{k=1}^{q} \beta_k x_{s_ir_ik}(t_i )\right\}}{\sum_{r\in R(t_i \mid \mathcal{H}_{t_{i}^-})} \exp\left\{\sum_{k=1}^{q} \beta_k x_{s_irk}(t_i)\right\}} \right).
\end{equation}
In its logarithmic form \eqref{eq:pl} assumes a concave behaviour, allowing the coefficients to be estimated via a Newton approach. 

\subsection{Case-control sampling of the risk-set and logit approximation}

The practical applicability of the partial likelihood is compromised by runtime complexities in the computation of its denominator, involving the risk-set $R(t| \mathcal{H}_{t^-})$ \citep{foucault_welles_dynamic_2014}. As already noted by \cite{butts_2008}, the risk-set typically grows quadraticly with the number of nodes in the network, making computations slow beyond a few hundred nodes. Even though the risk-set in our case consists only of alternative receivers, this still involves millions of patents, making the partial likelihood approach inaccessible for our problem. 

The solution suggested by \cite{vu_relational_2015} is to reduce computational complexity by applying nested case-control sampling on the risk set\citep{borgan_methods_1995}. The idea is to analyze all the observed events, i.e., citations or ``cases'', but only a small number of non-events, i.e., non-citations or ``controls.'' \cite{borgan_methods_1995} proved that maximum partial likelihood estimation with a nested case-control sampled risk-set yields a consistent and asymptotically estimator. This approach reduces the number of computing resources needed to build the risk set, however it still makes heavy use of computer memory. 

%For each event, a random subset of non-cited patents is sampled across the ones that are already present in the network at that time to compose the sampled risk-set. This approach reduces the number of computing resources needed to build the risk set, however it still makes heavy use of computer memory. 

Empirical evidence presented by \cite{lerner_reliability_2020} suggests that such asymptotic estimates can already be reliable with just one control per case. With a single control, the denominator in \eqref{eq:pl} is the sum of the rates for the cited patent with covariates $x_{s_ir_i}$ and a randomly sampled non-cited patent with covariates $x_{s_ir_i^*}$. Then sampled case-control version of the partial likelihood \eqref{eq:pl} is given as,
\begin{equation}\label{eq:pl2}
	\widetilde{PL}(\beta) = \prod_{i = 1}^n\left( \frac{\exp\left\{\sum_{k=1}^{q} \beta_k \left(x_{s_ir_ik}(t_i )-x_{s_ir_i^*k}(t_i )\right)\right\}}{1+ \exp\left\{\sum_{k=1}^{q} \beta_k \left(x_{s_ir_ik}(t_i )-x_{s_ir_i^*k}(t_i )\right)\right\}} \right),
\end{equation}
which is the likelihood of a logistic regression with only success and covariate levels $x_{srk}(t)-x_{sr^*k}(t)$. This approximation reduces the amount of memory needed to analyze the full set of observed citations, while the concavity of the logit approximation ensures the convergence of any Newtonian optimizers.  

%\cite{lerner_reliability_2020} empirically show that for REMs, reliable asymptotic estimates can be obtained even with only one control per case. In this latter case, the denominator in \eqref{eq:pl} would be just the sum of the rates for current cited patent and a sampled one that has not been cited. 

%becomes
%\begin{equation}\label{eq:logit}
%	\tilde{PL}(\beta) = \prod_{i =1}^n \Bigg( \frac{\exp\{\sum_{k=1}^{q} \beta_k [x_{srk}(t)-x_{r^*k}(t) ]\}}{1+\exp\{\sum_{k=1}^{q} \beta_k [x_{srk}(t)-x_{r^*k}(t)]\}} \Bigg),
%\end{equation}

%which is equivalent to 

\subsection{Basis expansion of covariates}

The basic REM assumes that the rate of an interaction between a sender $s$ and a receiver $r$ depends linearly on the covariates. Given the temporal complexity shown in Figure~\ref{fig:deposited-vs-cit}, it is unreasonable to imagine that this is violated in the patent citation network. Given the logistic interpretation of the case-control partial likelihood, we propose to extend the REM via a generalized additive framework \citep{hastie_gams_1986} by modelling single covariates via basis functions splines (B-splines) \citep{Schoenberg1946Contributions,Schoenberg1969interpolation}. 

B-splines are connected piece-wise polynomial functions of order $p$ defined over a grid of knots $u_0,u_1,\dots,u_m$, such that $u_{l-1}<u_{l}$, for $l=1,\dots,m$, on the parameter space that characterize the covariate $x_{srk}(t)$, for $k=1,\dots,q$.    The $j$-th B-spline basis function $B_{j, p}^k(x_{srk})$ ($j=1,\dots,d$) can be defined recursively \citep{DeBoor1972}, as
\begin{equation}\label{eq:bspline_formula}
B_{j,p}^k(x_{srk}) = \frac{x_{srk}(t )-u_j} {u_{j+p}-u_{j}}  B_{j,p-1}^k(x_{srk})	+ \frac{u_{j+p+1}-x_{srk}(t )}{u_{j+p+1}-u_{j+1}}  B_{j+1,p-1}^k(x_{srk}),
\end{equation}
where $d$ are the number of splines that represent the degrees of freedom of the B-spline transformation for covariate $x_{srk}(t)$ and where 
\begin{equation*}
	B_{j,0}^k (x_{srk})= 
	\begin{cases}
		1 & \text{if } u_j \leq x_{srk} < u_{j+1}\\
		0 & \text{otherwise}. 
	\end{cases}
\end{equation*}
In our modelling framework, we decided to place the knots evenly on the covariate support. The B-spline effect associated to a covariate $x_{srk}$, for $k= 1 \dots q$, is then a linear combination of $d$ coefficients with $d$ basis functions, i.e.,
\begin{equation*}
	f_k(x_{srk})=\sum_{j=1}^{d}\theta_{jk} B_{j,p}^k(x_{srk} ).
\end{equation*}
By substituting the basis expansion formulation in the hazard formulation in \ref{eq:rem}, the full model for the intensity function becomes 
\begin{equation}\label{eq:rem_gam}
	\lambda_{sr}(t \mid H_{t^-})= \lambda_0(t) e^{\sum_{k=1}^{q} f_k(x_{srk}(t))},
\end{equation}
where its partial likelihood approximation with one control is given as 
\begin{equation}\label{eq:logit_gam}
	\widetilde{PL}(\theta) = \prod_{i =1}^n \left[ 1+\exp\left\{- \sum_{k=1}^{q}\sum_{j=1}^d \theta_{jk}\left[B_{j,p}^k(x_{s_ir_ik}(t_i))- B_{j,p}^k(x_{s_ir_i^*k}(t_i))\right]\right\} \right]^{-1}.
\end{equation}

To smooth the estimated B-splines resulting from maximizing the partial likelihood in \ref{eq:logit_gam}, various penalization terms can be added. One reliable option is the use of P-splines \citep{eilers_flexible_1996}, especially when dealing with flexibility at the boundaries of the covariate support. However, in order to calculate the penalty, a considerable number of bases must be generated. Using large degrees of freedom translates to high memory usage, as each predictor generates two matrices of dimension $d$. As a result, over-parametrizing each predictor spline to provide smoothing can quickly exhaust the computer memory, making this procedure unsuitable for modelling large networks. In such situations, a cross-validation approach is preferred to select an appropriate number of basis functions, as memory constraints pose an upper limit on the number of degrees of freedom of the splines.

\subsection{Recovering baseline hazard}

The partial likelihood approach avoids modelling the baseline hazard. Although this brings significant benefits in estimating the B-splines, it also loses information about the underlying rate of the process. The advantage of formulating the REM as a Cox regression is that we can rely on the survival modelling literature to estimate the cumulative baseline hazard post-hoc.  We adapt the baseline estimator from the nested case-control sampling \citep{borgan_methods_1995} to estimate the underlying rate of the citation process. The adapted estimator for the cumulative baseline hazard is given as
\begin{equation}\label{eq:cumu_base}
	\widehat{\Lambda}_0(t) = \sum_{t_i < t} \left[ \exp \left\{  \sum_{k=1}^q f_k(x_{s_ir_ik}(t_i)) \right\} + \exp \left\{\sum_{k=1}^{q} f_k(x_{s_ir_i^*k}(t_i)) \right\} \right]^{-1} \frac{2}{n(t_i)},
\end{equation}
where $n(t_i)=|\mathcal{R}(t_i|\mathcal{H}_{t_i^-})|$ is the number of events at risk at $t_i$, for $i=1,\dots,n$.  A pointwise baseline hazard estimate can be calculated by taking differences between subsequent events of the cumulative hazard, i.e.,
\begin{equation}\label{eq:base}
	\hat{\lambda}_0(t_i) = \frac{\widehat{\Lambda}_0(t_i)-\widehat{\Lambda}_0(t_{i-1})}{t_i-t_{i-1}}, \quad \quad \text{for } i=1,\dots,n.
\end{equation}

\subsection{Parameter estimation using stochastic gradient descent}

While the case-control partial likelihood helps to reduce computational complexity, it is not be enough to overcome the optimization challenges presented by the size of the patent citation network.  Most machine-learning techniques use stochastic gradient descent methods to address large optimization challenges.  By separating the data stream into separate batches and adjusting the parameters after assessing each batch in succession optimization convergence can be achieved efficiently. As a result, even when working with large datasets, estimating the model parameters becomes manageable.

In this problem, we have opted for a stochastic gradient descent approach through the Adaptive Momentum (ADAM) optimizer \citep{kingma_adam_2017} to fit the partial likelihood. ADAM is a popular optimization algorithm that has been widely used in the field of machine learning. ADAM uses adaptive learning rates that depend on estimates of the first and second moments of the gradients of the observed batch. It maintains an exponentially decaying average of past gradients and squared gradients, which it then uses to calculate the update step for the model parameters. Let $\nabla PL(\theta)_b$ be the gradient evaluated on the partial likelihood on batch $b$. The ADAM routine updates the first and second moments according to the following routine:
\begin{align*}
	m^1_b &\leftarrow \xi_1 m^1_{b-1} + (1-\xi_1)\nabla \widetilde{PL}(\theta)b\\
	m^2_b & \leftarrow \xi_2 m^2_{b-1} + (1-\xi_2)\nabla \widetilde{PL}(\theta)_b,
\end{align*}
where $m^1$ and $m^2$ are the first and second moment gradients, respectively, and $\xi_1$ and $\xi_2$ are hyperparameters that control the importance of past information in updating the moments. The model parameters are then updated according to the following rule:
\begin{equation*}
	\theta_b \leftarrow \theta_{b-1} + \psi \frac{m^1_b}{m^2_b},
\end{equation*}
where $\psi$ is the learning rate that determines the magnitude of each parameter update.

The optimization procedure is repeated until the algorithm reaches the maximum point and the gradient becomes zero. At this stage, the method converges to a stationary distribution, indicating that the parameters have achieved a stable state where further parameter updates do not improve the model performance. However, it is important to note that the optimization process can be stopped earlier if the performance on a validation set starts deteriorating or if the maximum number of iterations is reached. Additionally, the ADAM algorithm has two built-in mechanisms to handle sparse gradients and correct bias in the estimates. 

First, in many real-world scenarios, gradients are often sparse, which means that only a small fraction of the parameters have non-zero gradients at any given time. In traditional gradient descent algorithms, these sparse gradients can result in slow convergence or even divergence. ADAM handles sparse gradients by incorporating a technique called moment correction, which adjusts the moment terms based on the frequency of non-zero gradients, which allows the optimizer to effectively use the sparse gradients.

Furthermore, the ADAM algorithm uses bias correction to account for the bias introduced in the first and second moments of the gradients. The bias correction is necessary because the moving averages of the gradients (the first and second moments) are initialized to zero and thus biased towards zero, especially at the beginning of the fitting process. To correct this bias, ADAM applies a correction term to the moving averages, which is proportional to the learning rate and inversely proportional to the number of iterations. This correction term ensures that the moving averages are unbiased estimates of the true first and second moments of the gradients.

Overall, ADAM has demonstrated its effectiveness as a reliable optimizer for various machine learning applications. As a result, STREAM is estimated efficiently for large number of observatins even with the addition of additive components described by B-splines.

\section{Modeling patent citations}\label{effects}

By using STREAM in evaluating patent citations, the key question we seek to answer is what are the underlying causes for a patent to be cited. The mechanisms that produce patent citation network can be endogenous and exogenous. Given the enormous size of the patent citation network, we will begin this section with a description of the model implementation. We then discuss the effects that we considered and how models including various effects have been compared. We complete the section with a discussion of the results we have found and their implications for the patent citation process. 

\subsection{Implementation}

Although the process of generating basis functions from events and estimating the coefficients can be tackled by well-optimized R algorithms like the \texttt{gam} function in the \texttt{mgcv} package \citep{wood2011mgcv}, it is unable to deal with 100 millions patent citations \citep{oancea2014integrating}. The R software memory management system struggles with large data objects, resulting in limitations to the practical applicability of routines, such as {\tt gam}. This complicates the estimation of the coefficients through the optimizers in \texttt{mgcv} as they would require a considerable amount of time to reach convergence. Spline basis expansions require the storage in memory of as many $n \times d$ matrices as there are covariates in the model. In fact, in the REM partial formulation \ref{eq:logit_gam}, this involve $2q$ matrices for both cases and controls. 

The model fitting problem can, therefore, be divided into two parts: (i) defining an efficient way to compute the basis function for millions of rows, and (ii) avoiding to generate matrices that exceed the available memory.

To tackle the first problem, we create a vectorized recursive algorithm that efficiently generates basis functions from millions of elements in a vector. Dividing the input data into batches is analogous as taking random samples from a larger population. The value associated with each observation following the basis function transformation is invariant to the position of the event in the observed set. Rather than applying the basis function transformation on the entire observed stream of the events, these can be computed directly on each batch when the gradient needs to be computed. This reduces memory usage at the expense of a small increase in computational costs.  The implementation relied on the Python suite PyTorch  \citep{pytorch}, which also provides access the computational benefits of Graphic Processor Units (GPUs) to scale matrix multiplications and gradient computations. The vectorized nature of Pytorch and the use of GPU computational power are particularly suited for the recursive algorithm, drastically reducing the computational time for generating B-splines. Then, by dividing the stream of data into different batches, we can efficiently estimate the coefficients by iteratively updating them with respect to the back-propagated gradients, computed using the negative log-partial likelihood from \ref{eq:logit_gam} as our loss metric. The code for STREAM can be found at \url{https://github.com/efm95/STREAM}.

\subsection{Potential drivers of patent citations}

In this section we describe the type of drivers we consider in the patent citation process. We divide the type of tested statistics into node effects, patent similarity effects, and time-varying effects, related to viral and saturation considerations of patent citations. Table~\ref{tab:effect_mechanisms} contains an overview of all the effects and their respective mechanisms.

\paragraph{Node effects.} Nodal effects refer to specific information about the cited patent, such as the publication year of the cited patent. A non-linear \emph{cited patent publication year} effect can uncover any tendencies where patents issued in specific years are being cited more consistently. This can potentially indicate a period of significant technological advancement or the emergence of a new technology. 

In addition, the time that has elapsed between the publication of the citing and cited patents is also a crucial factor in analyzing the patent citation network. This \emph{time lag} effect can provide insight into the tendency of patents to cite material that is closer in time, reflecting the current state of technological innovation. By counting the number of days between the citing patent issue date and the receiver publication date, we can model this \emph{time lag} effect and account for the time that has passed between the two nodes appearing in the network.

%With the advent of digitalization in patenting procedures, the number of issued patents has increased dramatically, leading to a rise in the number of backward citations \citep{kuhn_patent_2020}. 

Not all patents are equally important in terms of their connectivity. One hypothesis may be that if a particular patent summarizes a lot of older knowledge, it may attract more citations. This hypothesis is sometimes referred to as the \emph{cumulative process of knowledge creation} \citep{scotchmer1991standing}. To test for this hypothesis, we use the \emph{receiver outdegree} as a proxy for centrality. In this context, the outdegree of a patent represents the number of patents itself cites at time of its publication. 

\paragraph{Patent similarities effects.} Citations in patents arise from the assumption that there exists some technological similarity between the citing and the cited patent. However, technological relatedness is not a particularly tangible concept. Two distinct types of relatedness are often used to capture this idea.

The first type is a direct textual similarity between the citing and cited patents. Although there have been debates about the reliability of textual similarity as a measure of technological relatedness, recent studies \citep{kuhn_patent_2020, filippi-mazzola_drivers_2023} have shown that it plays an important role in patent citation networks, when combined with other metrics.  Following the same procedure described in \cite{filippi-mazzola_drivers_2023}, we calculate textual similarities of pair of patents through a pre-trained  Sentence-BERT neural network \citep{reimers-2019-sentence-bert}. It uses a vectorized loop to compute pairwise similarities of the abstracts of citing and cited patents.

%The first is a direct textual similarity between the citing and the cited patent. Although there has recently been a discussion on the reliability of textual similarity \citep{kuhn_patent_2020}, \cite{filippi-mazzola_drivers_2023} showed how the application of certain corrections in the model specification such as the time lag makes the textual similarity still a relevant component that needs to consider. Then, following the same procedure of \cite{filippi-mazzola_drivers_2023}, we computed the pairwise textual similarity across cited and citing patent abstracts by the means of a pre-trained Sentence-BERT model \citep{reimers-2019-sentence-bert}.

Another important measure of technological relatedness is the overlap in technology classes between the citing and the cited patents. Patent classification systems, such as the International Patent Classification (IPC) scheme, are designed to facilitate the search for related technologies by classifying patents into a systematic hierarchical structure. Deeper levels of the classification indicate higher levels of specificity in the technological field. However, analyzing patent classes presents several challenges. Patents often straddle various technological fields, and, as a result, may be allocated to multiple IPC classes. Furthermore, new IPC classes have been created, or existing ones have been merged or split since the creation of the USPTO \citep{younge_patent--patent_2015}, leading to a somewhat organic organization of technology classes. Despite these challenges, patent classes remain a crucial element in the patent-issuing process. To test the hypothesis that the assigned labels play a role in the citation process, we calculate the \emph{Jaccard similarity index} for the IPC classes of the cited and citing patents \citep{yan_measuring_2017}. The Jaccard index measures the similarity of the patent classes between two patents, taking into account the sub-class levels of the IPC classification. \cite{filippi-mazzola_drivers_2023} have shown that both the main component (Section) and the third component (sub-class) share similar importance in analyzing patent classes. 

\paragraph{Time-varying effects.} All the effects described above are fixed effects, corresponding to fixed characteristics of the dyad of citing and cited patent. We will also consider  two time-varying effects. Citation networks have distributional characteristics that are consistent with a viral process \citep{redner1998popular}. Patents that are popular, for whatever reason, may be more likely to draw more citations. We define for every patent  the \emph{cumulative number of citations received}. This is also known in network science as \emph{receiver indegree} or \emph{preferential attachment}. 

This popularity effect may experience saturation. For this reason, we also consider the \emph{time from the last event}, i.e., the last time the patent was cited. This variable captures how long it has been since the patent was last cited, and its influence on the rate of receiving a new citation.

Including time-varying effects into the model specification presents additional challenges in terms of computational efficiency. Specifically, updating the risk set for each observed event-time would require updating all the covariate levels of the time-varying components. Rather than uniformly sampling $x_{sr^*k}(t )$ from $R(t \mid \mathcal{H}_{t^-})$, we select a subset of candidates from the risk set, denoted by $\tilde{R}(t \mid \mathcal{H}_{t^-}) \subseteq R(t \mid \mathcal{H}_{t^-})$, such that for each event-time $t$, we sample $c$ potential receiver as candidates. For each candidate in $\tilde{R}(t \mid\mathcal{H}_{t^-})$, we update its relative time-varying effect at every observed time $t$. By storing these non-receiver candidates in memory, we significantly reduced the computational burden when creating the model matrix. It is important to note that the size of the subset $\tilde{R}(t \mid \mathcal{H}_{t^-})$ is determined by $c$, i.e. the number of potential non-event candidates for each observed event-time $t$. Although this approach may not capture all non-events, it provides an statistically consistent and computationally more efficient way of handling time-varying effects. 

Overall, incorporating time-varying effects in our model specification improves the accuracy and robustness of our analysis by accounting for the dynamic nature of patent citation behaviour over time. 

\begin{table}
	\caption{\label{tab:effect_mechanisms}Effects and their corresponding mechanisms. Where $r$ represents the cited patent, i.e. the receiver, $s$ represents the citing patent, i.e. the sender. Respectively, $t_s$ and $t_r$ represents the issue date for the sender for the receiver. \emph{Note}: $\mathbf{x}_{1r}$ and $\mathbf{x}_{1s}$ are the patents abstracts embedding vectors, while $\mathbf{x}_{2r}$ and $\mathbf{x}_{2s}$ are vectors of IPC patent classes.}
	\centering
	\begin{tabular}{*{3}{m{0.32\linewidth}}}
			\vspace{0.5cm} Effect               & 	\vspace{0.5cm} Mechanism                               & 	\vspace{0.5cm} Statistic   \\ \hline
		Receiver publication year&\includegraphics[scale=.75]{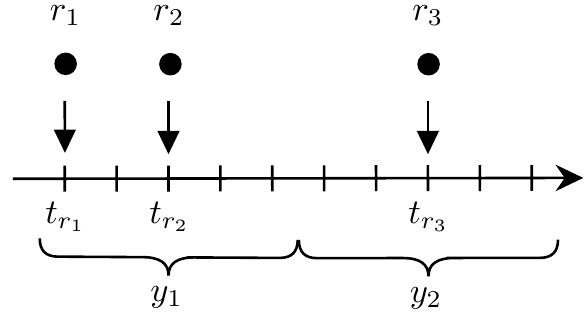} &$t_r$ \\
		\hline
		Time lag &\includegraphics[scale=.75]{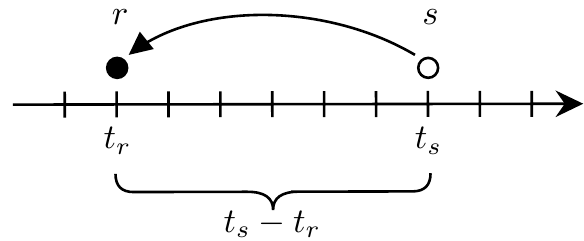} & $t_s-t_r$\\
		\hline
		Receiver outdegree &\includegraphics[scale=.75]{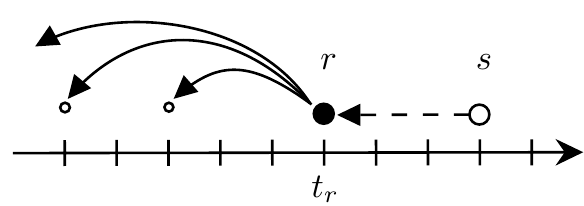} & $\sum_{r' \in R(t_r \mid \mathcal{H}_{t_{r}^{-}})} \mathbbm{1} _{\{(r,r',t_r)\}}$\\ 
		\hline
		Textual similarity & \includegraphics[scale=.75]{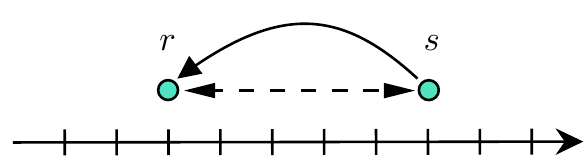} & $\frac{\mathbf{x}_{1r} \cdot \mathbf{x}_{1s}} {\|\mathbf{x}_{1r}\|  \|\mathbf{x}_{1s}\| }$\\ 
		\hline
		IPC relatedness & \includegraphics[scale=.75]{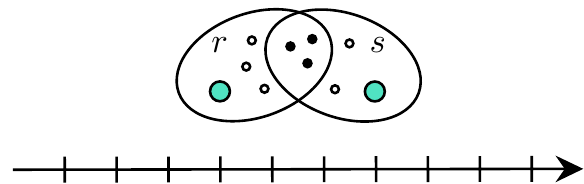} & ${{|\mathbf{x}_{2r}   \cap \mathbf{x}_{2s}|}\over{|\mathbf{x}_{2r}  \cup \mathbf{x}_{2s}|}} $ \\ 
		\hline 
		Cumulative citations received & \includegraphics[scale=.75]{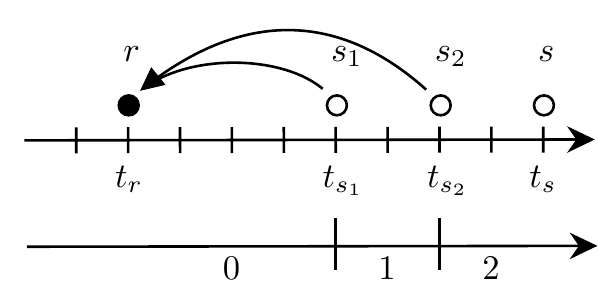}& $\sum_{t_i < t_s}\mathbbm{1}_{\{(s,r,t_i)\}}$\\
		\hline
		Time from last event & \includegraphics[scale=.75]{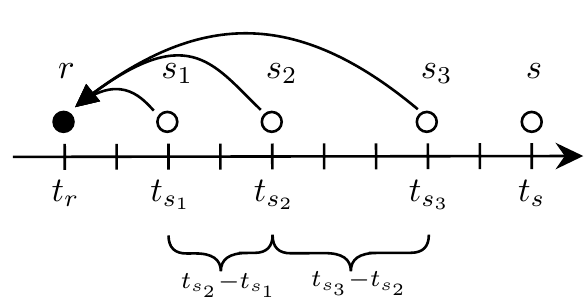}&  $\displaystyle \min _{t_i<t_s} \{(t_s-t_i) \mid \mathbbm{1}_{\{(s_i,r,t_i)=1\}}\}$\\ 
		\hline
	\end{tabular}
\end{table}

%The inclusion of time-varying effects to the model specification complicates the computation of the risk set because, for each observed event-time, the time-varying statistics for the non-events must be updated as if they had occurred at the event-time. When dealing with millions of events, updating a risk set the same size as the event set would result in a significant computational bottleneck. However, instead of  sampling uniformly $x_{sr^*}(t \mid H_{t^-})$ from $R(t \mid H_{t^-})$, we can define $\tilde{R}(t \mid H_{t^-}) \subseteq R(t \mid H_{t^-}) $ such that for each event-time $t$, we sample $k$ potential receiver as candidates.  Then, for every non-event candidate in $\tilde{R}(t \mid H_{t^-})$, we compute its relative time-varying at every observed time $t$. Thus, by saving these in memory these non-receiver candiates, we drastically reduce the computational complexity when creating the model matrix. The asymptotic property ot this approach is guaranteed by the fact that for big values of $k$, $\tilde{R}(t \mid H_{t^-}) \rightarrow R(t \mid H_{t^-}) $. 

\subsection{Model selection}

Various model formulations have been compared with each other on the available data. For simplicity, we grouped the statistics into \emph{nodal} (No), \emph{similarity} (Si), and \emph{time-varying} (Tv) effects, we sequentially add those groups of statistics to our model. On the left side of Table~\ref{tab:model_selection}, the estimated AIC and BIC values for each fitted model are reported. The results highlight the significant contribution that similarity statistics make to the model when they are included. It also suggests that the complete model is the best in describing the underlying data. Clearly, with so much data available, these relatively sensible statistics all significantly contribute to improving the fit of the relational event model. 

%\begin{table}
%	\caption{\label{tab:var_sel}Variable selection confronting AIC, BIC and log-Likelihood using three different model specifications. Effects have been divided into three groups: \emph{nodal} (No), \emph{similarity} (Si), and \emph{time-varying} (Tv).}
%	\centering
%	\begin{tabular}{l|c|c|c}
%		Effect group  & AIC           & BIC           & log LKL           \\ \hline
%		No 					& $101'708'032$ & $101'708'328$ & $-50'854'000$ \\
%		No + Si  		 & $22'777'864$  & $22'778'356$  & $-11'388'902$ \\
%		No + Si + Tv & $15'697'526$  & $15'698'214$  & $-7'848'721$  
%	\end{tabular}
%\end{table}

\begin{table}
	\caption{\label{tab:model_selection}\emph{Left}: Model  selection considerations using AIC, BIC and log-likelihood.  Effects have been divided into three groups: \emph{nodal} (No), \emph{similarity} (Si), and \emph{time-varying} (Tv). \emph{Right}: 6-fold cross-validation performed to compare three different batch sizes with varying degrees of freedom for the B-splines ranging from 4 to 20.}
	\centering
	\begin{tabular}{*{2}{m{0.49\linewidth}}}
		%\vspace{0.5cm}Variable selection               &  \vspace{0.5cm} Model selection \\ \hline
		\vspace{0.5cm}
		\begin{tabular}{l|l|l}
			Effect group  &AIC & BIC            \\ \hline
			No 					& $101'708'032$ & $101'708'328$ \\
			No + Si  		 & $22'777'864$  & $22'778'356$ \\
			No + Si + Tv & $15'697'526$&  $15'698'214$   \\
		\end{tabular}&\vspace{0.5cm}\includegraphics[scale=.5]{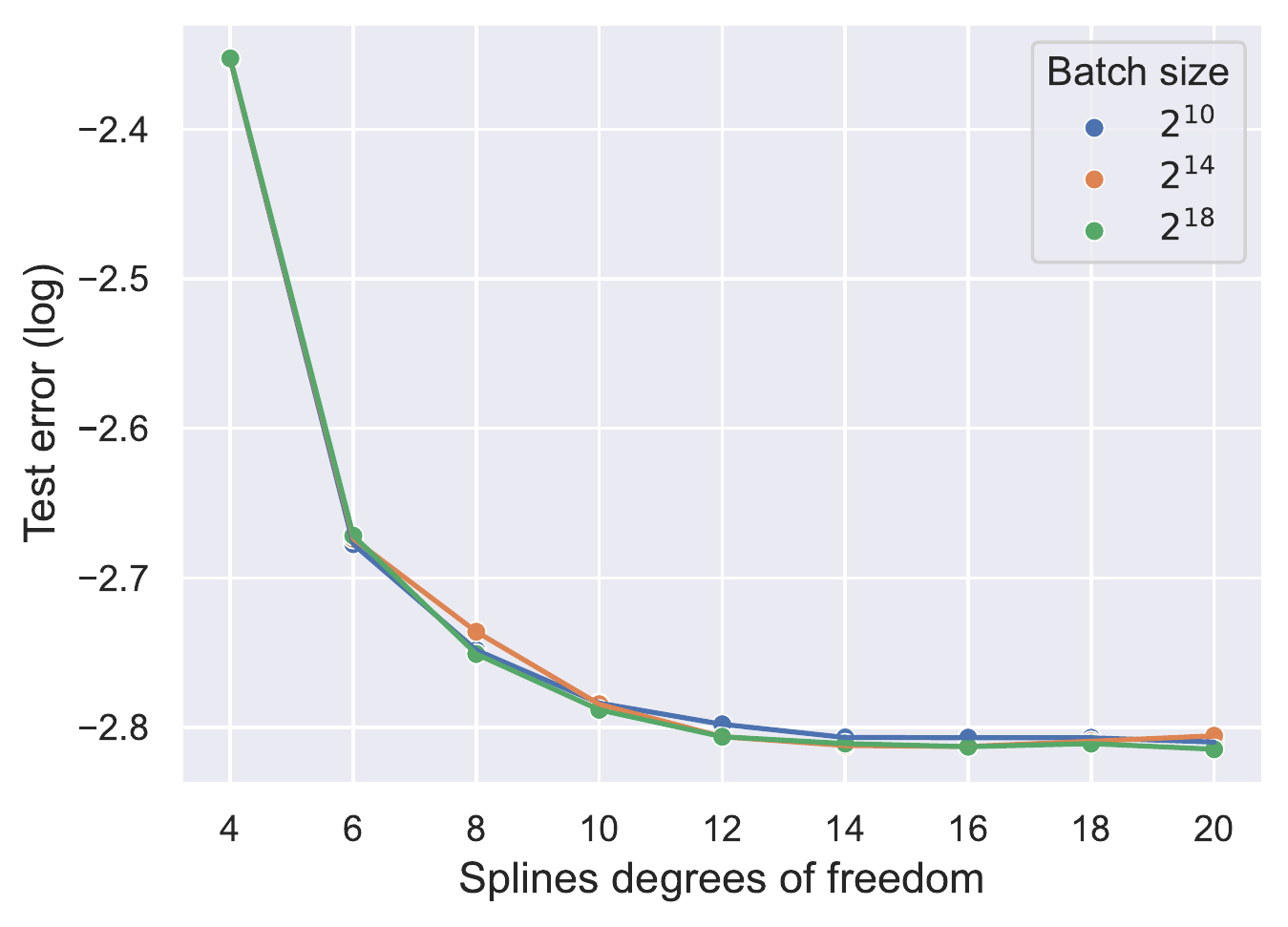} \\ 
	\end{tabular}
\end{table}

Furthermore, two hyperparameters need to be specified: batch sizes and the spline degrees of freedom.  Batch sizes refer to the number of events used in each batch during the model fitting process. We tested three different batch sizes: $2^{10}$, $2^{14}$ and $2^{18}$. The choice of batch size affects the trade-off between computational efficiency and  accuracy of model fit. Smaller batch size induce noisier gradients in the parameter updates, but are computationally more efficient. 

The degrees of freedom refers to the flexibility of the nonparametric model used to estimate the non-linear effects. We tested degrees of freedom ranging from 4 to 20 to find the optimal level of flexibility. A lower degree of freedom may result in a less flexible model that underfits the data, while a higher degree of freedom may result in a model that is too flexible and, potentially, overfits the data.

To determine the best values for these hyperparameters, we use a 6-fold cross-validation approach, where we test these three different batch sizes while changing the degrees of freedom from 4 to 20.  As a test-error metric, we evaluate the negative log-likelihood for each held-out set in the cross-validation. These values were then re-scaled by the size of their validation set for comparison purposes. 

%\begin{figure}
%	\centering
%	\includegraphics[width=\textwidth]{model_selection_log.pdf}
%	\caption{\label{fig:model_selection}6-fold cross-validation approach used to compare three different batch sizes with different degrees of freedom for the splines ranging from 4 to 20.}
%\end{figure}

The results are shown on the right of Table~\ref{tab:model_selection}. This figure shows the average test error for three different batch sizes ($2^{10}$, $2^{14}$, and $2^{18}$) and degrees of freedom ranging from 4 to 20. The results indicate that the average test error is statistically equivalent for the three batch sizes, suggesting that increasing the batch size beyond $2^{10}$ does not improve model fit. However, we observe that increasing the batch size from $2^{10}$ to $2^{14}$ results in only a marginal increase in the time to convergence, indicating that there is only a slight increase in computational demand between these two sizes. Therefore, we conclude that the optimal batch size is $2^{14}$ as it provides the best trade-off between stability and time-to-convergence. Furthermore, the plot shows that the test errors stabilize after the 12th degree. Therefore, we select $12$ for the B-spline degrees of freedom. 

\subsection{Interpretation of results on USPTO patent citation data 1976-2022}

\begin{figure}
	\centering
	\includegraphics[width=\textwidth]{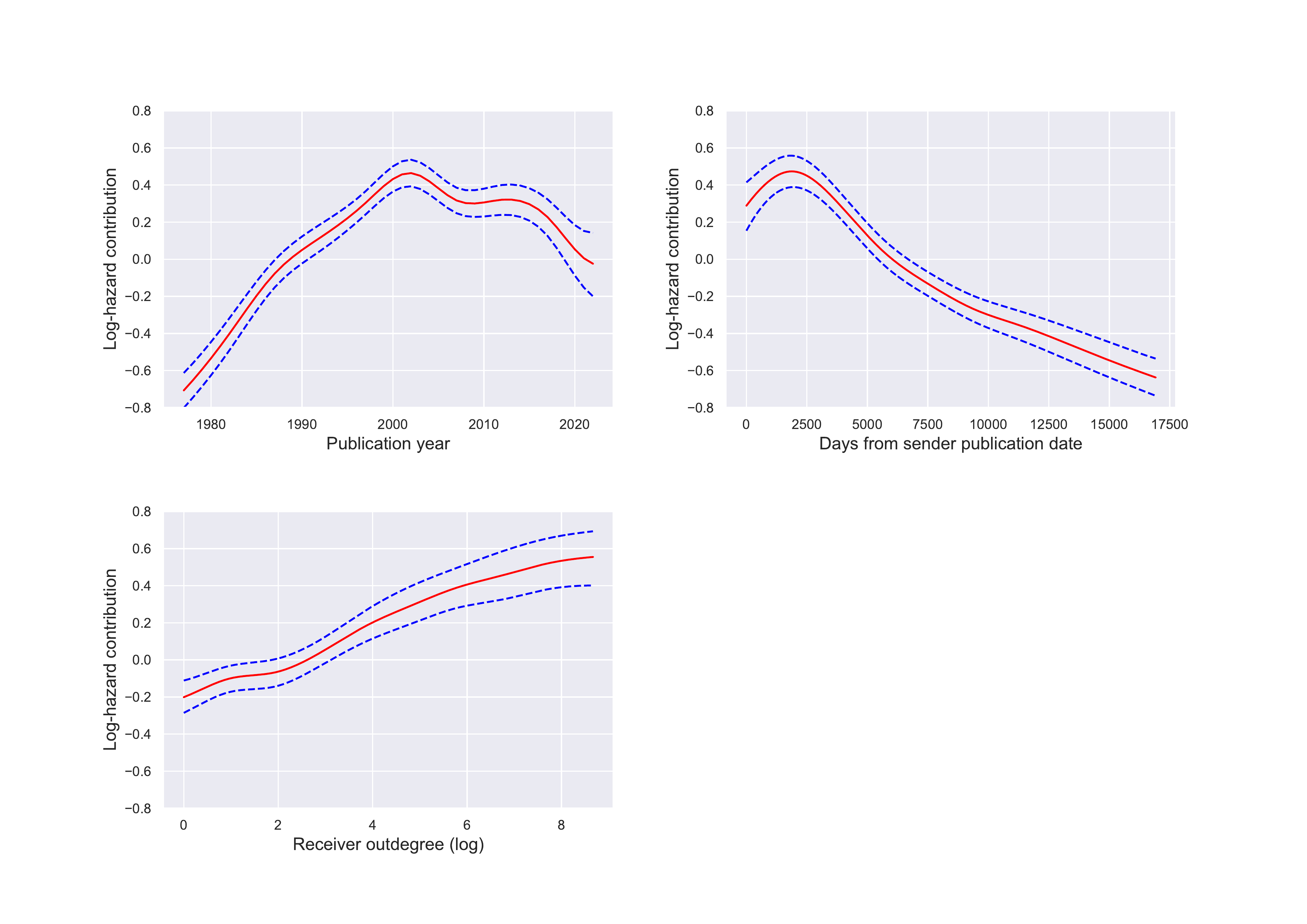}
	\caption{\label{fig:no} Splines associated to the \emph{nodal effects}. Up-left: \emph{receiver publication year}. Up-right: \emph{time lag}. Down-left: \emph{receiver outdegree} in log terms.}
\end{figure}

\begin{figure}[tb]
	\centering
	\includegraphics[width=\textwidth]{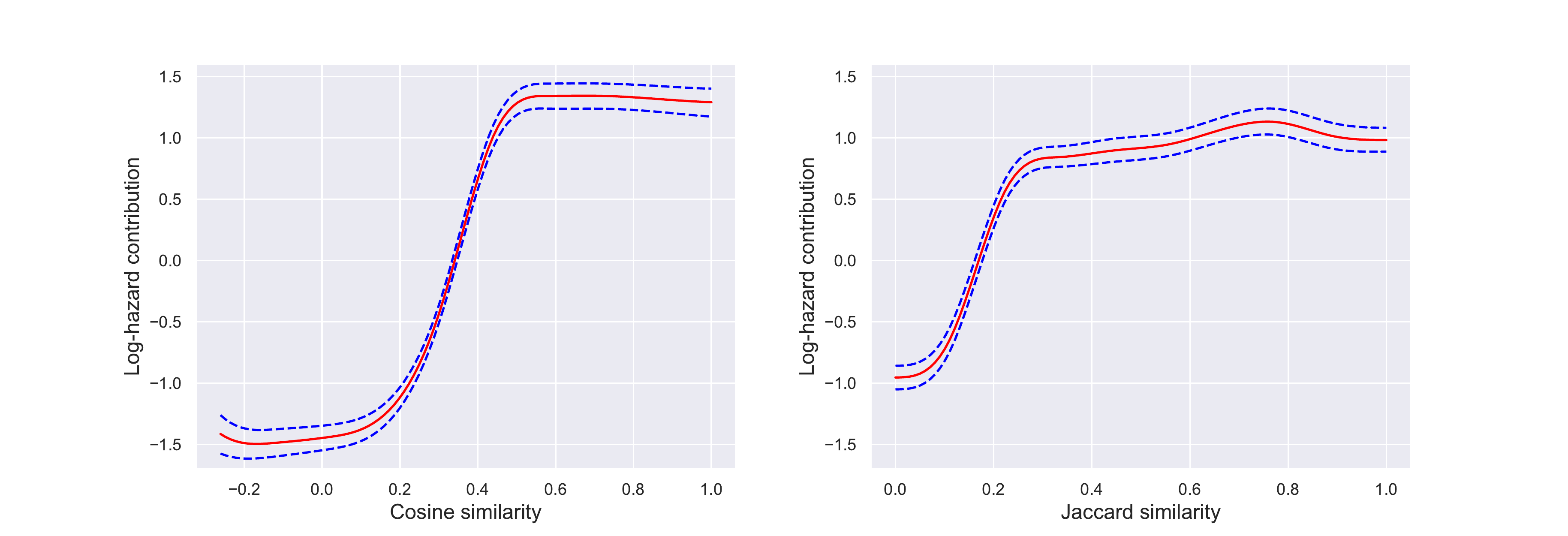}
	\caption{\label{fig:sim}Splines associated to the \emph{similarity effects}. Left: \emph{textual similarity}. Right: \emph{IPC relatedness.}}
\end{figure}

\begin{figure}[tb]
	\centering
	\includegraphics[width=\textwidth]{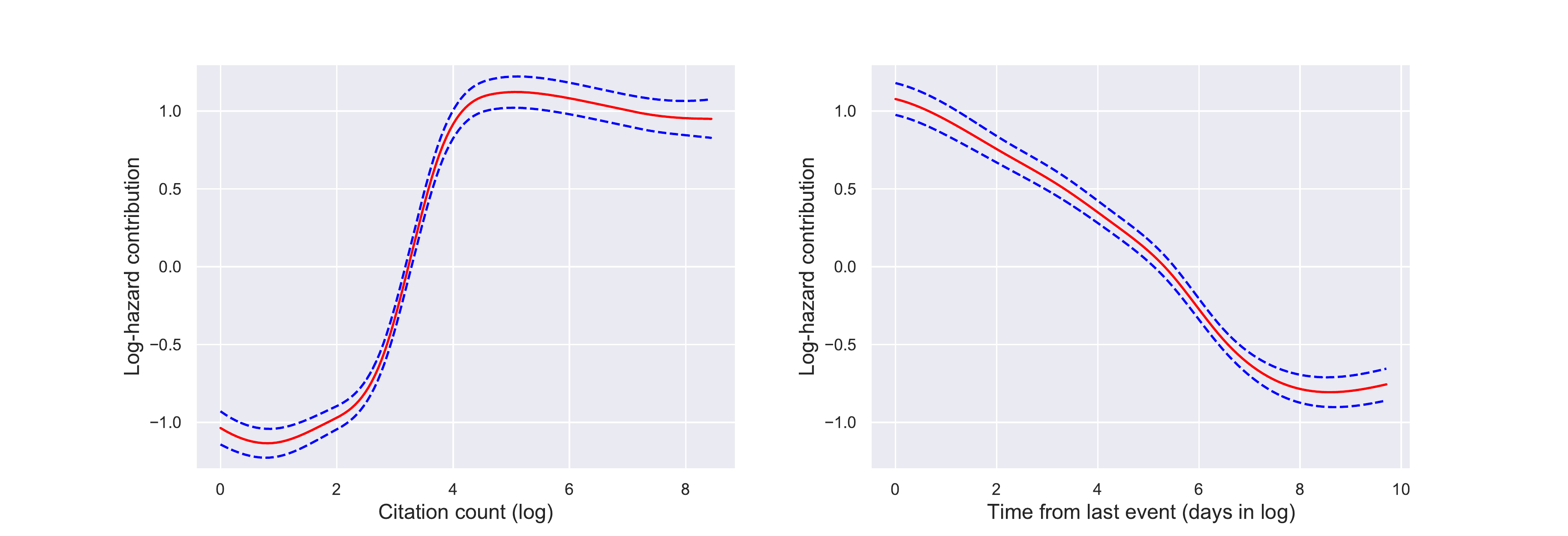}
	\caption{\label{fig:tv}Splines associated with the \emph{time-varying effects}. Left: \emph{cumulative citations received} in log scale. Right: \emph{time from last event} in days in log scale.}
\end{figure}

The stochastic component of the optimizing method introduces some randomness into the estimation of the model parameters. The approach is reliable, as it converges quite rapidly. Figures~\ref{fig:no}, \ref{fig:sim} and \ref{fig:tv} show the fitted splines with 12 degrees of freedom, including their uncertainty estimates, based on 50 resamplings. The y-axes indicate the log-hazard contribution to the citation rate of an individual patent. An increase by 0.7 on this scale indicates a doubling of the citation rate.

%THIS PART SHOULD BE IN THE METHODOLOGY, NOT IN THE RESULTS.
% IN FACT, WE NEED TO TALK ABOUT THIS:
%To asses the uncertainty of the estimated parameters, repetitions of the fitting procedure are needed. Most famous software packages that fit splines in their models solve this issue by repeating the model fit through bootstrapped versions of the model matrix \citep{wood2011mgcv}. However, in the context of a temporal relational event network, where each event is either unique or the result of an important endogenous mechanism of the network, having parametric or nonparametric repetitions of the same event could bring to important alterations in the model matrix that could lead to biases in the fitting routine. In the STREAM, we repeat the fitting process multiple times, shuffling the positions of the model matrix such that the procedure starts every time from a different point. To bring more strength to our results, after 50 repetitions, $\tilde{R}(t \mid \mathcal{H}_{t^-})$ was resampled. This procedure was repeated 10 times.  

\paragraph{Node effects.} %From the nodal statistics fitted splines in Figure~\ref{fig:no}, we can identify particular characteristics that increase the likelihood of a patent being cited. 

One remarkable result can be seen in the \emph{receiver publication year} curve in Figure~\ref{fig:no}. Contrary to the continuous increase in the patent depositing and patent citation process, the rate with which an individual patent gets cited possess a distinct peak. The peak occurs shortly after the year 2000. This means that, after accounting for all other effects, patents that were published around 2000 are, individually, attracting more citations that at any other period from 1976 to 2022. In fact, patents from around 2000 tend to attract 60\% more citations than those published around 2022, and more than 3 times more citations than those published in 1976.  While this study is limited to a macro-level network analysis, we hypothesize that several key technical breakthroughs may have occurred around 2000. \cite{park2023papers} also reported the recent decline in the disruptiveness of patents. However, in contrast to their findings, we find clear evidence for monotone increasing innovation from 1976 before peaking around the year 2000. It may be that by failing to take into account the growing patent network, their initial decline is an artifact. 

%At the end of the parameter space, there is a downward trend in the log-hazard contribution. This could be due to patents being too young to be cited. This trend is also supported by the confidence bands, which show that the interval widens around the year 2019. Interestingly, patents published at the beginning of the observed period have a negative log-hazard contribution. This could be a countereffect caused by the temporal boundaries of the network. However, it is worth noting that the curve assumes a positive trend only during the year 1990, where the boundaries effect should be less significant. Furthermore, following the findings of \cite{filippi-mazzola_drivers_2023}, the introduction of a time-lag effect could help mitigate these network boundary effects. 
%Taken together, these observations provide further evidence of the occurrence of important technological innovations during the peak periods. The \emph{receiver publication year} curve sheds light on the temporal dynamics of the citation process, indicating that certain periods are more conducive to citation. 

The \emph{temporal lag} spline indicates at which time in the future patents are most likely to be cited. The curve shows that there is a peak around day 2500. This indicates the presence of a sweet spot of approximately 7 years after the original publication of the patent where citations are most likely to arise. It is important to note that this temporal lag effect could be influenced by various factors such as the pace of technological development, the lifespan of technology, and the overall trends in the field. Indeed, this effect provides valuable insights into the timing of patent publications and their impact on the citation network. By identifying this sweet spot where citations are more likely to arise, inventors can strategically plan their patent filing and publication strategies to increase their chances of being cited and recognized in the field. Furthermore, the inclusion of \emph{temporal lag} into the model deals with the boundary problem, as it account for the fact that recently published patents are unlikely to have gathered a significant number of citations.

The \emph{receiver outdegree} effect shows how being at the center of the network structure leads to a bigger log-hazard contribution. The fitted spline suggests that patents that cite a lot of other patents are more likely to be cited themselves. This finding highlights the importance of citing all relevant patents in one's patent application, as it makes a patent more visible and accessible to other inventors, increasing the likelihood of being cited. This result is consistent with previous research that has emphasized the importance of network position in predicting innovation outcomes \citep{uzzi1997social}. Furthermore, this finding has practical implications for policymakers and inventors who may wish to increase the likelihood of their patents being cited. By fostering collaboration and networking opportunities, inventors can improve their chances of being connected to other inventors and increase their outdegree, thus increasing the visibility of their work.

\paragraph{Similarity effects.} The curves for both \emph{textual similarity} and \emph{IPC relatedness} shown in Figure~\ref{fig:sim} demonstrate the significance of technological similarity between the citing and cited patents. It highlights how patents that are more closely related are more likely to cite each other. The \emph{textual similarity} curve indicates a stronger tendency towards citing patents that share linguistically similar abstracts. The \emph{IPC relatedness} curve, on the other hand, indicates that patents that share even a limited number of technology classes have a higher probability of being cited.

Furthermore, the weight placed on the \emph{textual similarity} effect is noteworthy. Compared to patents that share a linguistic similarity less than 0.2, patents that share a similarity larger than 0.5 are 20 times more likely to cite each other. While the \emph{IPC relatedness} effect is not as strong as the \emph{textual similarity} effect, it still increases the citation rate by more than 7 times, beween patents that share at least 0.3 IPC classification on the Jaccard scale.  

These findings confirm previous studies \citep{jaffe_innovation_2002}. Despite the structural changes over time in the technological similarity across cited and citing patents \citep{kuhn_patent_2020,whalen_patent_2020}, patents with greater technological similarity remain more likely to cite each other.

\paragraph{Time varying effects.} The two time-varying effects in Figure~\ref{fig:tv} demonstrate the dynamic nature of patent citations. The \emph{cumulative citation count} effect reveals how the number of citations a patent has received so far influences its likelihood of being cited in the future. This effect is particularly notable as the log-hazard contribution shows a rapid increase after receiving more than 20 citations, indicating a positive feedback loop where the more citations a patent receives, the more likely it is to receive additional citations. This snowball effect is a crucial factor in determining the significance of a patent within the network, and it underscores the importance of early recognition and citation of relevant breakthroughs.

On the other hand, the \emph{time from last event} effect highlights the inverse relationship between the time interval from the last citation and the likelihood of receiving subsequent citations. As the time interval grows longer, the probability of receiving additional citations decreases. This effect is shown by the steady decrease in log-hazard contribution. This trend underlines the importance of continuous and timely recognition of relevant patents to maintain their significance and relevance within the citation network.

It is worth noting that these two effects work in together to shape the dynamic nature of patent citations within the network. The snowball effect of the \emph{cumulative citation count} effect can counteract the decay caused by the \emph{time from last event} effect, but only up to a certain point. Eventually, even the most significant patents will fade in relevance if they are not consistently recognized and cited within the network. %This underscores the importance of a timely and accurate recognition of relevant patents to drive innovation and technological progress.

\subsection{Estimated baseline hazard}

\begin{figure}
 	\centering
 	\includegraphics[width=0.7\textwidth]{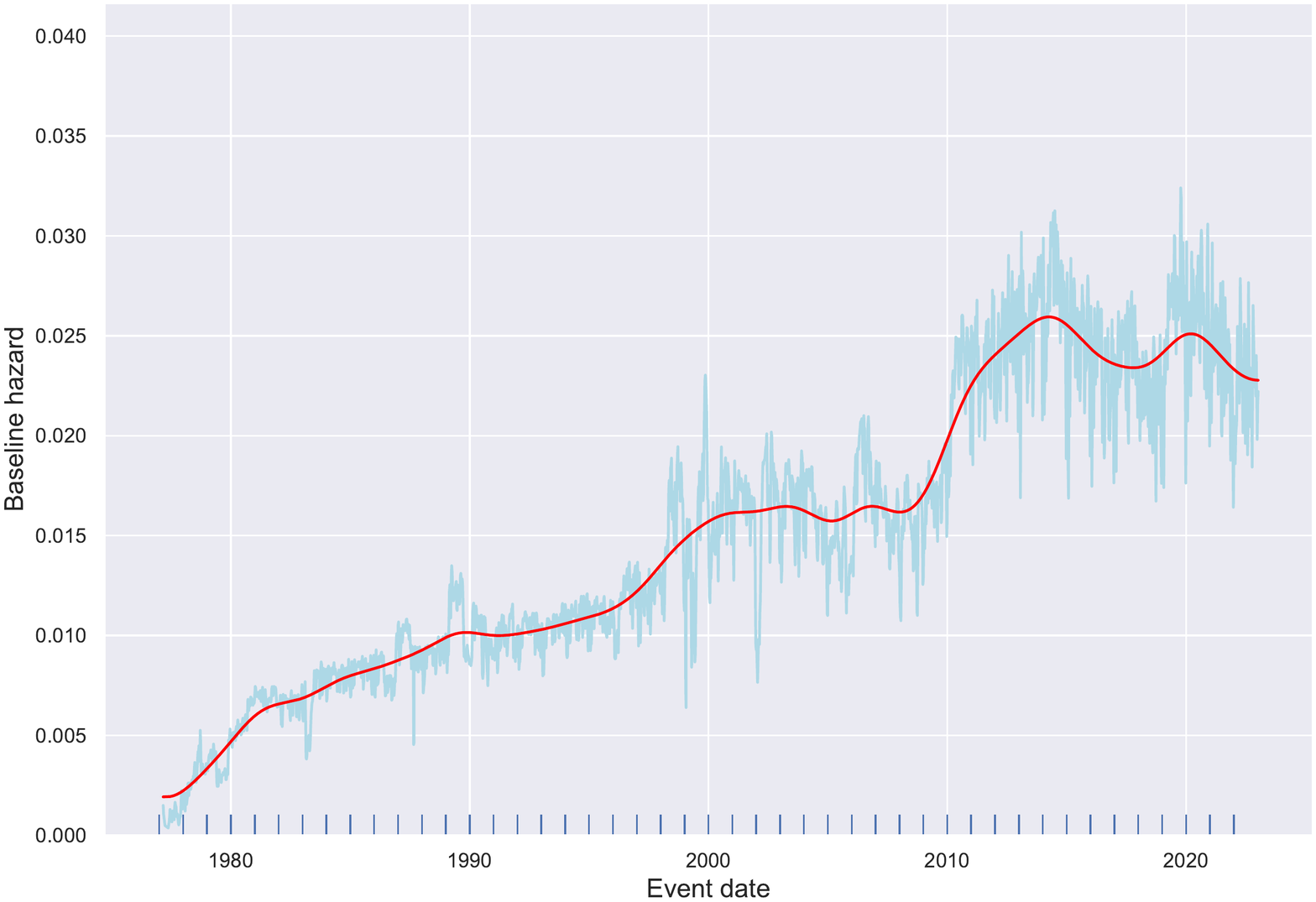}
 	\caption{\label{fig:base}Baseline hazard estimated through the adapted estimator in (\ref{eq:base}) from \cite{borgan_methods_1995}. The red line is the application of a Gaussian filter to smooth the resulting estimate and capture the general trend of the baseline.}
 \end{figure}

To analyze the overall pattern of hazard rates over time, we estimated the baseline hazard by differentiating the adapted version of the \cite{borgan_methods_1995} estimator presented in (\ref{eq:base}), using the average coefficients obtained from the repeated fits of the STREAM. To capture the general trend and present a clearer picture of the base hazard, we applied a Gaussian filter to the estimated baseline. Figure~\ref{fig:base} shows the estimated baseline hazard, which provides a visual representation of the overall pattern of the hazard rates over the observed period.
 
As anticipated, the curve demonstrates that the baseline rate of being cited increases over time. The general increasing trend of the curve indicates that patents become more susceptible to being cited as time goes on, which may be attributed to the accumulation of knowledge and technological advancement over time. This observation is consistent with the general expectation that the probability of being cited will grow over time due to the increasing size and complexity of the patent network. Moreover, this result underscores the importance of considering the temporal dimension when analyzing citation patterns and provides valuable insights into the dynamics of knowledge diffusion in patent systems.
 
Furthermore, the estimated baseline reveals an interesting trend in the patent citation network. Specifically, we note the sudden increase in the baseline hazard in the year 2010. It suggests that there may have been significant changes in the patent citation process during that time frame. One possible explanation for this observed increase is the legal changes in the applicant's duty of disclosure that took place in 2010. As reported by \cite{kuhn_information_2010}, these legal changes led to a drastic increase in the number and scope of cited references in patent documents. Consequently, more citations were included that were further afield from the citing patent, resulting in a generally higher rate of patent citations.

\section{Conclusions}

Relational event models are a sophisticated and effective approach for analyzing complex patterns in temporal network event data. In this study, we applied this framework to patent analysis for identifying the drivers of patent citations. The limitations of current fitting procedures for REMs make them, however, unsuitable for inferring large event networks. To overcome these limitations, we developed a Stochastic Gradient Relational Event Additive Model (STREAM) that combines non-linear modelling with machine learning techniques. By applying STREAM to a network of patent citations spanning from 1976 to the end of 2022 with over 8 million patents and over 100 million citations, we were able to identify patterns that affect the patent citation rate.

Our findings offer several interesting insights. While some effects are straightforward, others reveal peculiar patterns that require further investigation. For instance, we found that patents from around the year 2000 have been much more influential than from any other period. This suggest that there must have been several important technological innovations in those year. This information could be useful for policymakers, industry professionals, and researchers who want to identify important areas of innovation and potentially invest resources in those areas. %In this regard, 

There are several ways the analysis can be extended. Further research is required to assess which areas of technology have been innovating more and how this has developed through time. We could also consider a more sophisticated approach to incorporate the possible time decay of some effects. It would be interesting to evaluate the behaviour of, e.g., the \emph{textual similarity} curve in Figure~\ref{fig:sim} over the observed period, particularly in light of recent discussions on changes in the generative process of patent citations \citep{kuhn_patent_2020,filippi-mazzola_drivers_2023}. However, such studies would require careful consideration with respect to the underlying changes in the legal patent framework. Approaches like the ones proposed by \cite{juozaitiene_non-parametric_2022} could be further investigated to be applied to the STREAM to asses temporal decay of predictors. 

The STREAM approach could be further extended beyond citation networks to other types of dynamic networks. Although REMs pose computational challenges, STREAM provides a promising solution to overcome these challenges and open REMs to other large event networks. 

\section*{Acknowledgments}
This work was supported by funding from the Swiss National Science Foundation (SNSF grant 192549).

\bibliographystyle{apalike}
\bibliography{biblio}

\end{document}